# Photonic chip-based Reconfigurable Radar Compound Jamming Signal Generator

Senyu Zhang[1,2], Wei Luo[1,2], Shuang Zheng[1,2,*], Yunlong Li[1,2], Yiqiang Ou[1,2], Qi Tian[1,2], Yifei Chong[1,2], Zihang Yang[1,2], Li Shen[1], Minming Zhang[1,2,*]

[1]*School of Optical and Electronic Information and Wuhan National Laboratory for Optoelectronics, Wuhan, Hubei 430074, China*

[2]*National Engineering Research Center for Next Generation Internet Access System, Wuhan, Hubei 430074, China*

\* *email:* [zshust@hust.edu.cn](zshust@hust.edu.cn); [mmz@hust.edu.cn](mmz@hust.edu.cn)

## Abstract

Microwave photonics (MWP) serves as a powerful bridge between the radio-frequency and optical worlds, unlocking unprecedented bandwidth and speed for critical information systems. Recently, MWP-based radar jamming has demonstrated significant potential in overcoming electronic bottlenecks; however, existing solutions rely heavily on bulky discrete components and generate deterministic symmetric waveforms that are vulnerable to advanced counter-countermeasures. Here, we break this status quo by demonstrating the first monolithic photonic chip-based reconfigurable radar compound jamming signal generator. Implemented on a compact 2×5 mm² silicon photonic footprint, our system features high integration and electrical reconfigurability, supporting agile switching among four distinct jamming modes. Crucially, we introduce a composite jamming mechanism that exploits time-frequency coupling to break the inherent symmetry of traditional deceptive jamming, generating controllable asymmetric false-target clusters. The on-chip system exhibits an operating bandwidth exceeding 20 GHz and is capable of generating over 20 high-fidelity false targets simultaneously. This work establishes a fully functional platform combining broad bandwidth, multi-mode flexibility, and asymmetric deception, paving the way for miniaturized, intelligent photonic solutions in next-generation electronic warfare.

## Introduction

Dominance over the electromagnetic spectrum has become a decisive factor in modern warfare, where radar systems serve as the primary sensors for battlefield awareness and precision guidance. As radar technologies evolve toward ultra-wide bandwidths and cognitive capabilities, the challenge for electronic warfare (EW) systems has escalated from simple noise suppression to the generation of high-fidelity, multidimensional deceptive jamming[1–6]. Conventional electronic countermeasures, predominantly relying on digital radio frequency memory (DRFM)[7–10], utilize analog-to-digital and digital-to-analog converters (ADCs/DACs) to store and replicate radar signals. However, these electronic systems are currently hitting a fundamental bottleneck: constrained by the sampling rates and resolution of converters, they struggle to process the wideband, high-frequency signals prevalent in modern radar operations, severely limiting their instantaneous bandwidth and response latency.

Microwave photonics (MWP) has emerged as a revolutionary bridge between the radio-frequency and optical domains, offering a pathway to circumvent these electronic limitations by leveraging the inherent advantages of optics in bandwidth, speed, and electromagnetic immunity[11–14]. Over the past decade, diverse photonic jamming architectures have been proposed to replicate radar echoes[15–29]. Early efforts utilized optical storage loops based on long fibers to achieve high-fidelity signal memory[15–17], but these systems often suffered from optical self-oscillation and noise accumulation. To introduce Doppler velocity deception, acousto-optic modulators (AOMs) were integrated into the loops[18–21], yet the limited bandwidth of AOMs restricted the tuning range of frequency shifts. More recently, to confront sophisticated radar waveforms, advanced architectures have been developed to enhance jamming complexity. For instance, parallel modulation schemes have been employed to synthesize compound jamming signals[26], while multi-path optical delay lines[28] and cascaded modulator structures for non-uniform interrupted sampling[29] have been proposed to significantly densify false targets and disrupt their regularity. Despite these functional advancements, a significant gap remains between laboratory demonstrations and practical application. Most reported systems—including the aforementioned advanced architectures—are constructed using discrete fiber-optic components, resulting in bulky footprints and complex fiber interconnections that are highly susceptible to environmental vibrations. To overcome these integration barriers, the field has witnessed a paradigm shift toward chip-scale solutions, exemplified by the recent demonstrations of general-purpose programmable photonic processors[30–34]. While these pioneering works mark a significant step toward integrated reconfigurable systems, applying such monolithic platforms to specific electronic warfare tasks requires further architectural innovation.

Crucially, however, achieving integration solves only half the problem. Beyond the hardware form factor, a fundamental physical vulnerability pervades existing photonic jamming approaches: the deterministic symmetry of the generated false targets. In classic schemes like frequency-shifting jamming (FSJ) and interrupted-sampling repeater jamming (ISRJ), the false targets are typically distributed symmetrically around the real target with predictable amplitude envelopes. This symmetric signature acts as a recognizable fingerprint, allowing advanced radar signal processing algorithms to easily identify and filter out the jamming signals[35–40]. While recent efforts have attempted to generate asymmetric patterns, they remain confined to complex discrete systems[28,29]. Therefore, creating a robust integrated platform that not only capitalizes on the stability of monolithic integration but also possesses the unique reconfigurability to break this inherent symmetry has become a critical imperative for next-generation EW systems.

To address these challenges, we report the first demonstration of a reconfigurable radar compound jamming system implemented on a monolithic silicon photonic integrated chip. We have designed and fabricated a device within a compact 2×5 mm² footprint that monolithically integrates a high-speed modulator array and a photodetector, offering an operating bandwidth exceeding 20 GHz. Through flexible electrical programming, the chip allows agile switching among four distinct jamming modes without altering the hardware architecture. Most significantly, to overcome the symmetry limitation, we introduce a novel gated-modulation jamming (GMJ) mechanism that exploits the time-frequency coupling of chirped radar signals to synthesize controllable asymmetric false-target clusters. This work elevates complex microwave photonic jamming from discrete platforms to the chip scale, presenting an integrated solution to the urgent demands of modern electronic warfare for miniaturized, multi-functional, and intelligent countermeasures.

# Result

**Architecture and principle of the monolithic jamming processor**

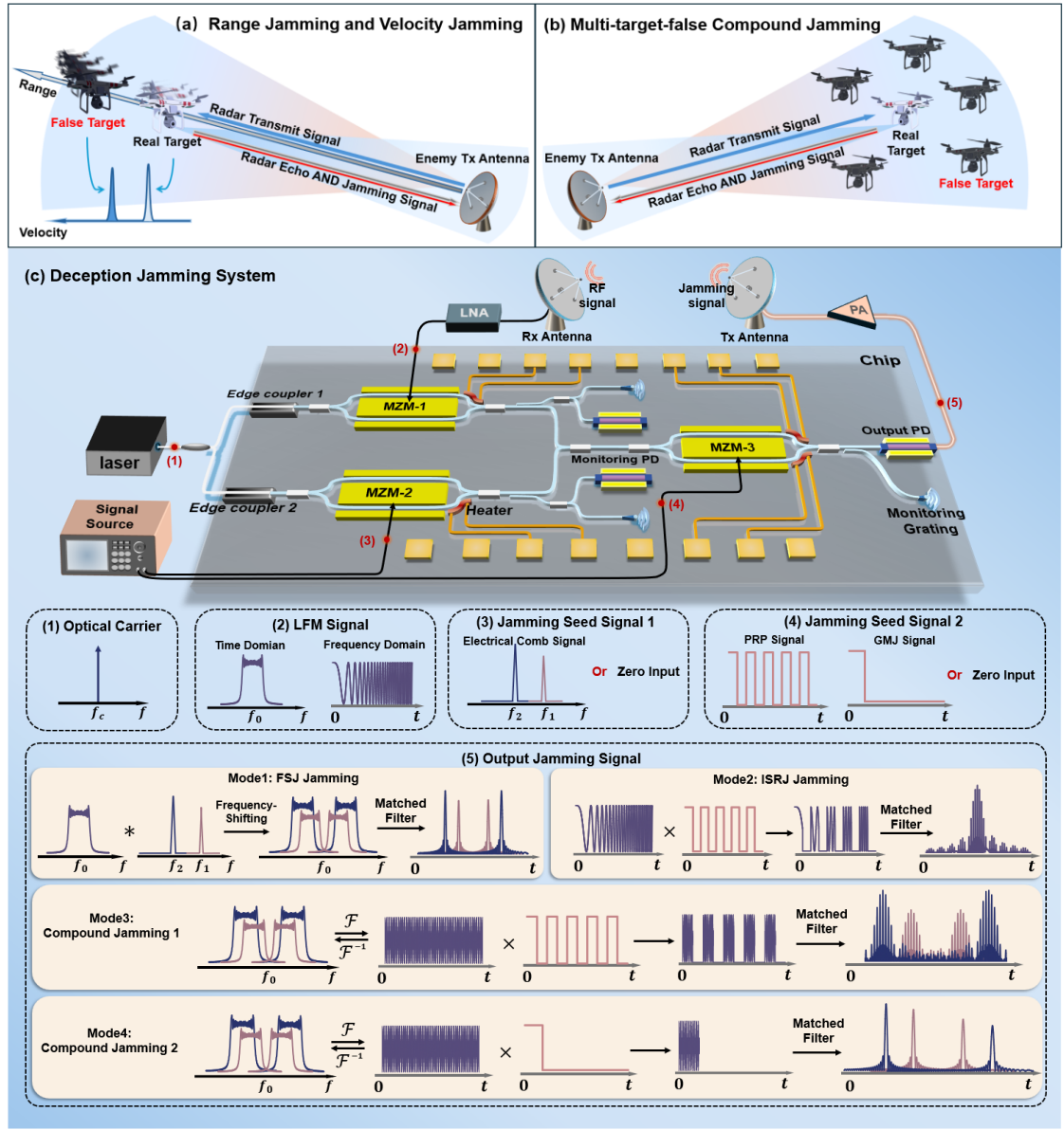

**Fig. 1 | Concept, architecture, and operating modes of the monolithic radar jamming processor. a, b** Operational scenarios illustrating the evolution of electronic warfare requirements: **a** Fundamental jamming in range and velocity dimensions to generate deceptive false targets. **b** Sophisticated multi-target compound jamming to create dense and complex false-target clusters for saturating cognitive radar resources. **c** Schematic of the monolithic silicon photonic integrated circuit (PIC). **d** Principles of the four reconfigurable jamming modes enabled by different seed signal combinations: FSJ via multi-tone comb modulation; ISRJ via periodic pulse sampling; Compound Jamming (FSJ+ISRJ) for enhanced target density; and Asymmetric Compound Jamming (FSJ+GMJ), which exploits time-frequency coupling to break the symmetry of false targets for advanced range-gate pull-off.

We propose a chip-scale photonic solution designed to meet the escalating demands of modern electronic warfare. The conceptual framework and evolution of jamming requirements are illustrated in Fig. 1. While current jamming systems effectively address fundamental dimensions such as range and velocity deception (Fig. 1a), the trend in radar countermeasures is shifting towards the generation of dense, multi-target compound jamming environments to saturate cognitive radar resources (Fig. 1b). To realize these sophisticated capabilities within a minimal footprint, we designed and fabricated a silicon photonic integrated circuit (PIC) (Fig. 1c). This chip, occupying a compact area of only 2×5 mm², monolithically integrates a high-speed signal processing core comprising three cascaded Mach-Zehnder modulators (MZM-1, MZM-2, MZM-3) and a high-bandwidth photodetector (PD). Crucially, to ensure stable operation and precise calibration, on-chip monitoring units including optical monitoring gratings and monitoring PDs are embedded alongside the processing core. This dense integration eliminates the phase instabilities inherent in discrete fiber-optic systems, while the use of an external laser source ensures high-power, low-noise optical carrier injection.

The core innovation of this platform lies in its flexible signal loading mechanism and reconfigurability. As shown in Fig. 1d, the intercepted radar RF signal is loaded onto MZM-1, serving as the base waveform. Subsequently, MZM-2 and MZM-3 are driven by specific jamming seed signals, such as electrical combs, periodic rectangular pulse (PRP) signal, or gated waveforms, to imprint deceptive features onto the radar signal. By selecting different combinations of these seed signals, the chip can be electrically programmed to switch agilely among four distinct jamming functions. FSJ and ISRJ provide baseline capabilities for range and velocity deception, creating symmetric false targets. To enhance complexity, compound jamming (FSJ+ISRJ) combines these effects to form a dense interference texture. Most significantly, we introduce asymmetric compound jamming (FSJ+GMJ), which exploits time-frequency coupling to break the inherent symmetry of traditional deceptive jamming, generating controllable asymmetric false-target clusters for advanced range-gate pull-off.

**Chip fabrication and characterization**

To translate the conceptual architecture into a tangible platform, the photonic integrated chip was fabricated using a commercial CMOS-compatible silicon photonics process (CUMEC), as shown in Fig. 2b. The chip is packaged into a compact optoelectronic module (Fig. 2a) via wire bonding to a custom-designed high-speed printed circuit board (PCB). Optical coupling is achieved through a high-precision fiber array (Fig. 2c) with a coupling loss of approximately 2.5 dB/facet, connecting the external laser source to the on-chip waveguide network.

The core processing units are designed for high-bandwidth operation. The modulators employ a single-drive series push-pull (SPP) traveling-wave electrode architecture (Fig. 2d). To ensure efficient broadband signal injection, the RF transmission lines on the PCB are optimized, yielding a 3-dB bandwidth exceeding 40 GHz (Fig. 2e). Detailed characterization of the MZM reveals an electro-optic bandwidth of 20 GHz at −2 V bias. Crucially for high-fidelity jamming signal generation, the thermal tuning curve (Fig. 2f) demonstrates a high extinction ratio exceeding 40 dB, with a measured half-wave voltage ($V\pi$) of approximately 3.9 V. Signal reconversion is performed by an on-chip high-speed germanium-on-silicon (Ge-on-Si) photodetector (Fig. 2g). The end-to-end characterization of the device shows a 3-dB electro-optic bandwidth of approximately 31.3 GHz, providing a robust physical foundation for generating wideband jamming signals in the K-band and beyond.

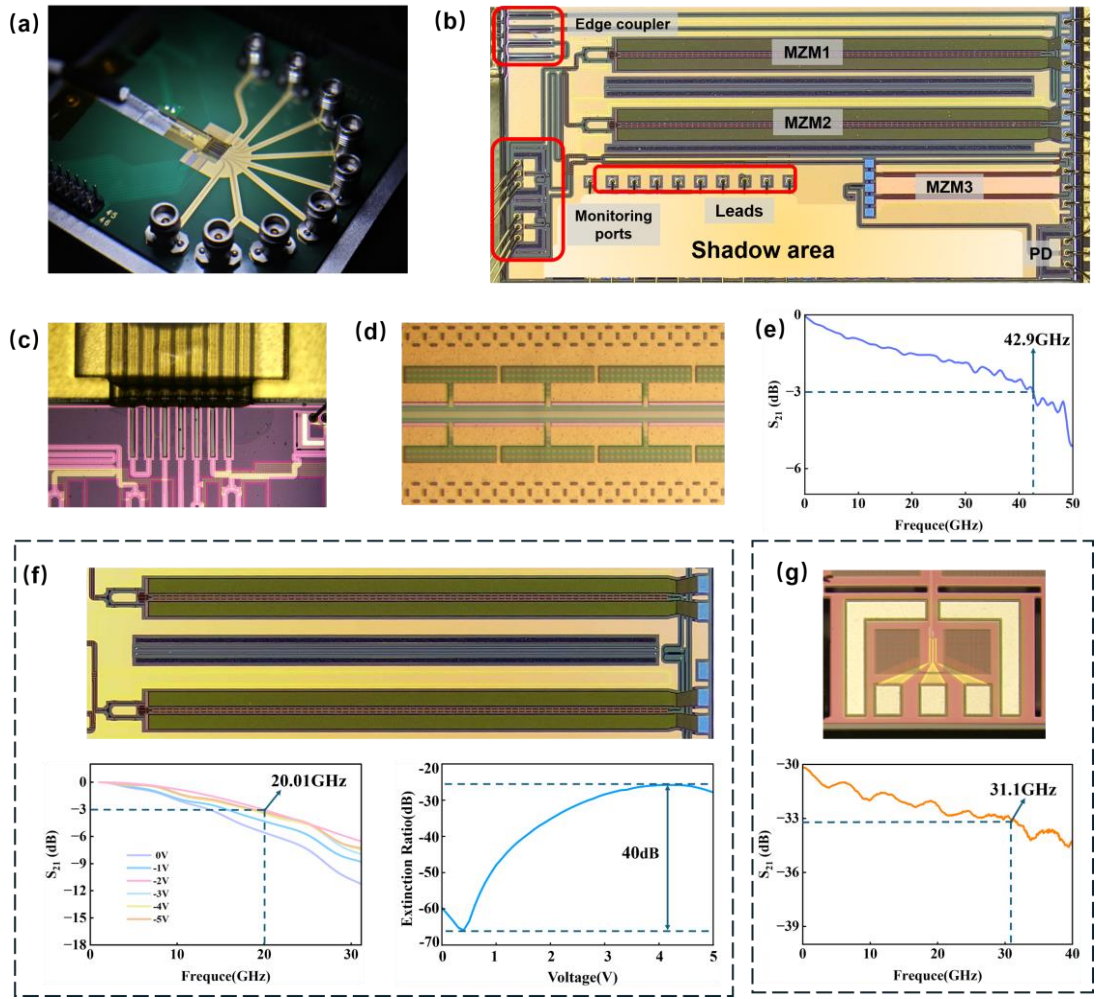

**Fig. 2 | Device implementation and characterization. a** Photograph of the packaged optoelectronic jamming module. **b** Micrograph of the fabricated silicon photonic chip. **c** High-precision fiber array (FA) coupled to the chip edge. **d** Microscope image showing the traveling-wave electrodes of the single-drive series push-pull Mach-Zehnder modulator (SPP-MZM). **e** Measured S21 parameter of the high-speed PCB RF port, showing a bandwidth exceeding 40 GHz. **f** MZM characterization: electro-optic bandwidth measurements at different bias voltages (top) and thermal tuning curve (bottom) indicating a half-wave voltage (Vπ) of ~3.9 V and an extinction ratio >40 dB. **g** Micrograph and measured electro-optic (E-O) S21 response of the on-chip germanium-on-silicon high-speed photodetector.

**Generation of symmetric false targets via FSJ and ISRJ**

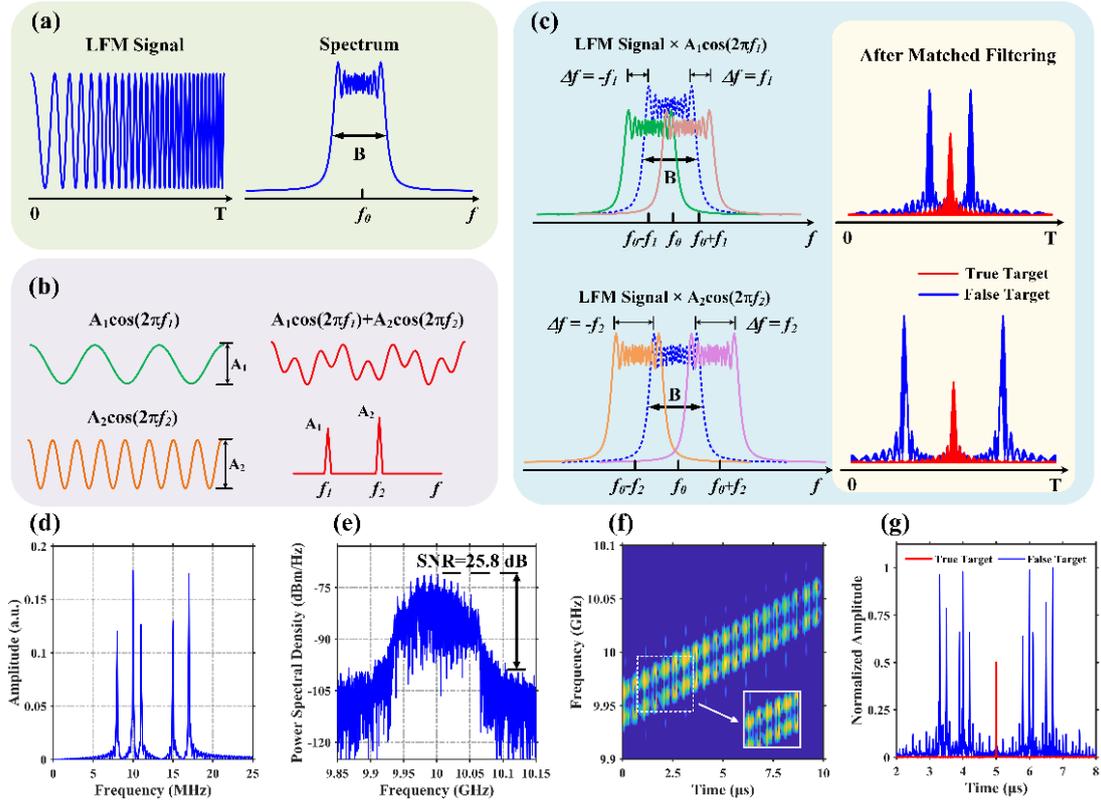

**Fig. 3 | Principle and experimental validation of on-chip FSJ. a** Schematic of the time-domain and frequency-domain representations of the LFM radar signal. **b** Synthesis principle of the electrical comb signal for frequency shifting. **c** Schematic of the FSJ principle: time-domain multiplication (equivalent to frequency-domain convolution) leads to spectral shifting and symmetric false targets. **d** Measured electrical spectrum of the five-tone comb seed signal. **e** Power spectral density (PSD) of the generated FSJ jamming signal. **f** Time-frequency analysis of the jamming signal. **g** False-target distribution after matched filtering (the real target is marked in red).

Following device characterization, we experimentally validated the system's baseline jamming capabilities. Figure 3 and Figure 4 illustrate the operational principles and experimental results for the two foundational modes: FSJ and ISRJ. The FSJ mode operates on the principle of time-domain multiplication between the intercepted radar signal and a multi-tone electrical comb (Fig. 3a–c). This operation is mathematically equivalent to a spectral convolution, which replicates the original radar spectrum at new frequency positions determined by the comb spacing. In our experiment, the chip was configured for FSJ mode by loading an intercepted LFM signal (10 GHz center frequency, 100 MHz bandwidth) onto MZM-1 and a five-tone electrical jamming seed signal (composed of sine waves at 8, 10, 11, 15, and 17 MHz) onto MZM-2. To ensure the signal passed through the subsequent stage without alteration, MZM-3 was biased at its Maximum Transmission Point (MATP) with no electrical drive.

The experimental results are detailed in Fig. 3d–g. The electrical spectrum of the seed signal is shown in Fig. 3d. As depicted in the power spectral density (PSD) (Fig. 3e) and time-frequency analysis (Fig. 3f), the generated jamming signal exhibits a high signal-to-noise ratio (SNR) of 25.8 dB and clearly reproduces the chirp characteristics. The matched filtering result (Fig. 3g) demonstrates the successful generation of ten discrete false targets surrounding the real target (marked in red). A key advantage of this on-chip FSJ mode is its precise controllability: by adjusting the frequencies and relative amplitudes of the electrical comb components, the positions and power levels of the generated false targets can be deterministically manipulated.

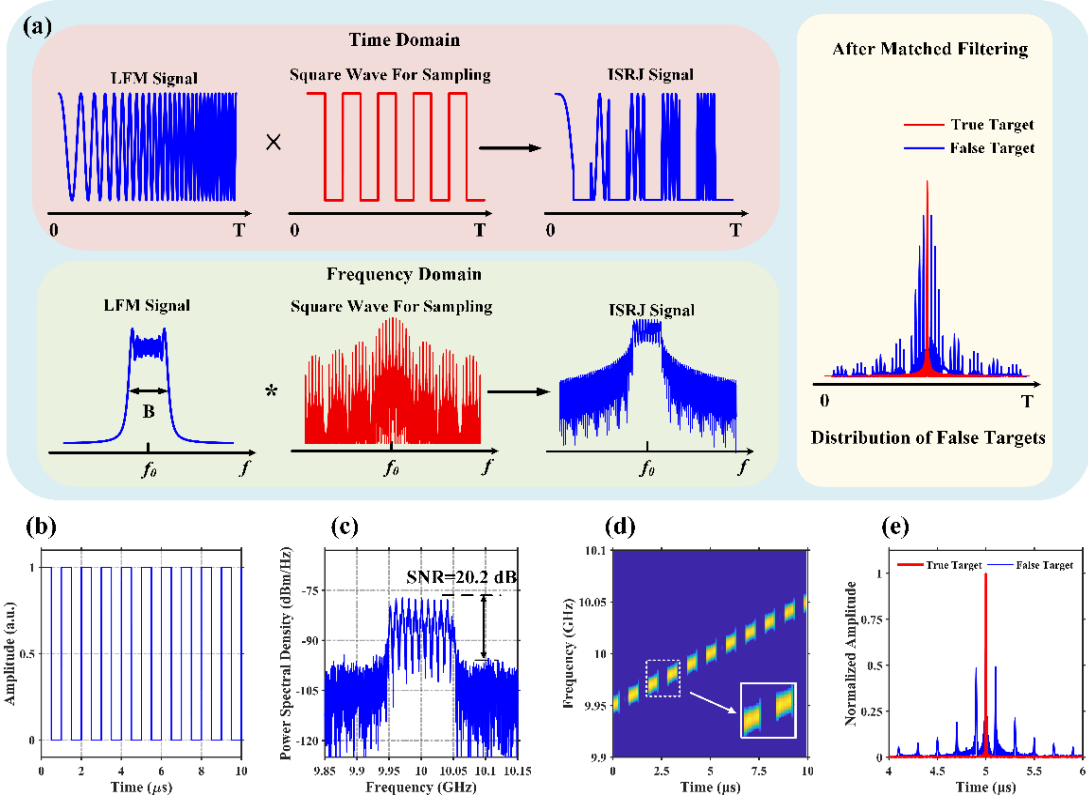

**Fig. 4 | Principle and experimental validation of on-chip ISRJ. a** Schematic of the ISRJ principle: the radar signal is periodically sampled by a square wave in the time domain, which is mathematically equivalent to a convolution with a harmonic series in the frequency domain. **b** Measured electrical spectrum of the square-wave jamming seed signal (1 MHz frequency, 50% duty cycle). **c** Power spectral density (PSD) of the generated ISRJ jamming signal. **d** Time-frequency analysis showing the periodic signal interruptions. **e** Dense false-target distribution after matched filtering, exhibiting a characteristic Sinc-function envelope.

Alternatively, the system can be reconfigured for ISRJ mode by utilizing the third modulator (MZM-3) as a high-speed optical switch. In this configuration, MZM-2 is biased at its Minimum Transmission Point (MITP) with no RF input to effectively suppress unwanted interference paths. By driving MZM-3 with a periodic rectangular pulse train (e.g., 1 MHz frequency, 50% duty cycle), the intercepted radar signal undergoes periodic time-domain "chopping." The core principle, illustrated in Fig. 4a, relies on the mathematical equivalence between time-domain sampling and frequency-domain convolution. Since the spectrum of a periodic rectangular pulse train consists of discrete harmonics enveloped by a Sinc function, this modulation effectively multicasts the radar signal onto these harmonic frequencies. Consequently, ISRJ generates a series of discrete false targets, mimicking a superposition of multiple frequency-shifted replicas with varying amplitudes. The spacing of these false targets is deterministically governed by the sampling frequency, while their overall amplitude envelope follows the Sinc function defined by the sampling duty cycle. The experimental results shown in Fig. 4b–e validate this mechanism, demonstrating the generation of a dense cluster of false targets with the predicted spectral and temporal characteristics. Beyond standard periodic sampling, this on-chip ISRJ mode offers significant flexibility: the distribution of false targets can be tailored by tuning the frequency and duty cycle of the square-wave signal. Furthermore, the system supports the loading of non-periodic pulse trains to synthesize even more complex and unpredictable jamming patterns.

However, despite the high fidelity and flexibility demonstrated in these experiments, a critical vulnerability remains. As observed in the matched filtering results (Fig. 3g and Fig. 4c), the false target distributions exhibit highly regular structures—strict symmetry in FSJ and a deterministic Sinc-function envelope in ISRJ. These distinctive signatures act as recognizable fingerprints, rendering the jamming signals susceptible to identification and suppression by advanced counter-countermeasure algorithms.

**Composite jamming: enhanced complexity via cascaded modulation**

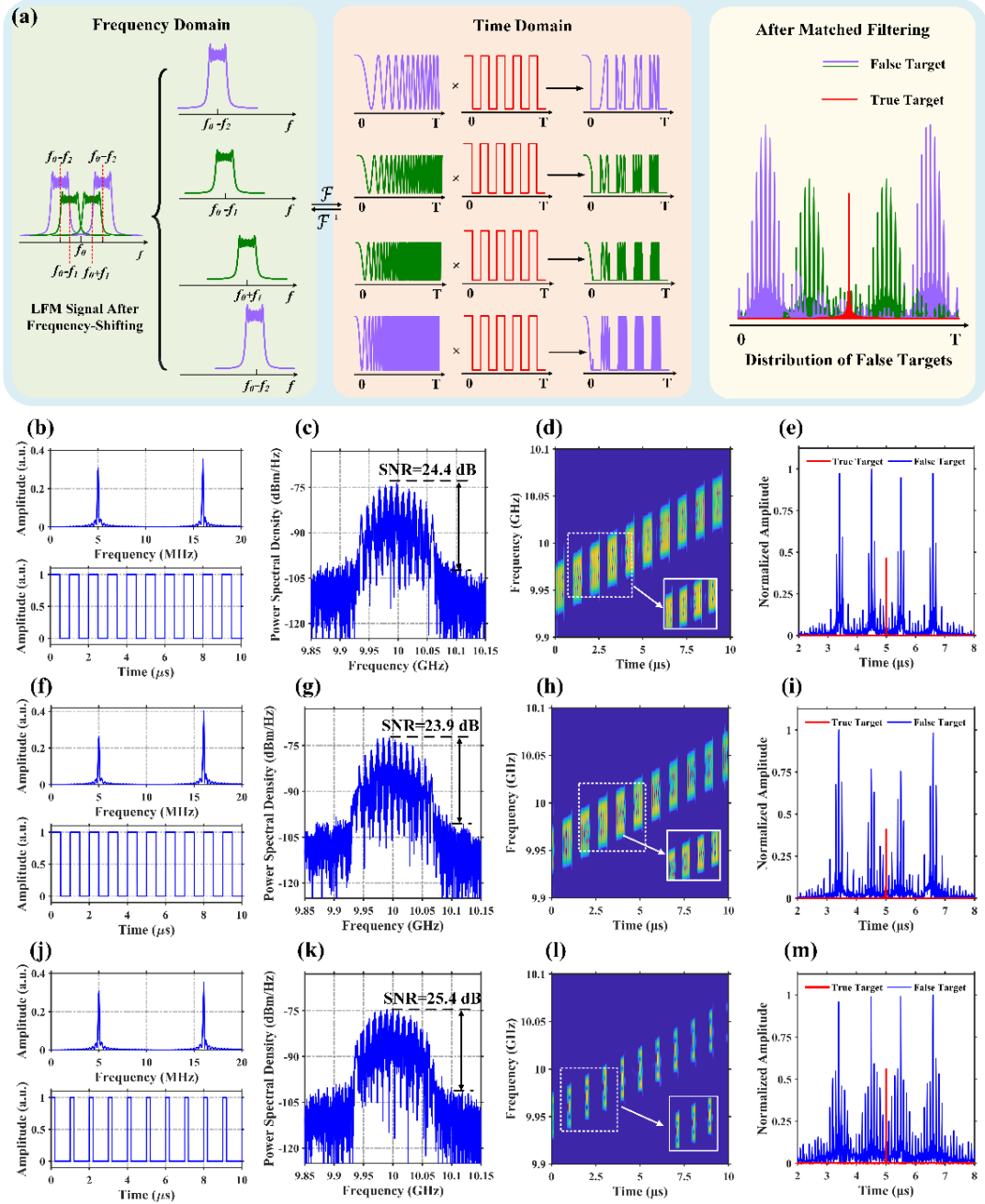

**Fig. 5 | Principle and experimental validation of Compound Jamming (FSJ+ISRJ). a** Schematic of the composite jamming principle: the radar signal undergoes sequential frequency shifting and interrupted sampling, leading to a

multiplicative expansion of false targets. **b–e** Experimental results for Set 1 (FSJ: 5 & 16 MHz, amplitude ratio 1:1; ISRJ: 1 MHz, 50% duty cycle). **b** Electrical spectrum of seed signals. **c** Power spectral density. **d** Time-frequency analysis. **e** False-target distribution after matched filtering. **f–i** Experimental results for Set 2 (FSJ amplitude ratio adjusted to 0.7:1; other parameters same as Set 1). **j–m** Experimental results for Set 3 (ISRJ duty cycle reduced to 20%; other parameters same as Set 1). Note: In all cases, the generated false targets exhibit symmetric distributions (real target marked in red).

To overcome the limitations of single-mode jamming, we exploited the cascaded architecture of the photonic chip to realize a composite jamming mode (FSJ+ISRJ). As illustrated in Fig. 5a, this mode operates by sequentially applying frequency shifting and interrupted sampling to the radar signal. Mathematically, this process corresponds to a convolution of the radar spectrum with both the multi-tone comb and the sampling harmonics, resulting in a multiplicative expansion of false targets. Specifically, each primary false target generated by FSJ serves as a center for a secondary cluster of targets generated by ISRJ, significantly increasing the density and complexity of the jamming environment. In the experimental implementation (Fig. 5b–m), the LFM signal and the frequency-shifting electrical comb were applied to MZM-1 and MZM-2, respectively, both biased at the Minimum Transmission Point (MITP) to perform the first stage of optical double-sideband suppressed-carrier modulation. The combined optical signal was then routed to MZM-3, which operated at the Quadrature Transmission Point (QTP) and was driven by a periodic rectangular pulse (PRP) train to execute the second stage of interrupted sampling.

Figure 5 presents the experimental results under three distinct parameter configurations, demonstrating the high reconfigurability of this mode. In the first set (Fig. 5b–e), we utilized a two-tone electrical comb (5 MHz and 16 MHz, amplitude ratio 1:1) and a 1-MHz, 50%-duty-cycle sampling wave. The matched filtering result (Fig. 5e) reveals that the four primary false targets from FSJ evolved into distinct clusters, each comprising three sub-targets, increasing the total count to 12 with a high SNR of 24.4 dB. In the second set (Fig. 5f–i), by adjusting the amplitude ratio of the frequency-shifting signal to 0.7:1, we observed a corresponding variation in the cluster amplitudes, confirming the system's controllability at the "cluster" level. Finally, in the third set (Fig. 5j–m), reducing the sampling duty cycle from 50% to 20% significantly increased the number of sub-targets within each cluster. This configuration successfully generated approximately 20 effective false targets, substantially increasing the jamming density to further obscure the real target. While this composite mode markedly increases scene complexity, a higher-order symmetry persists. As evident in the results, both the macroscopic layout of the clusters and the microscopic structure within each cluster remain symmetric. This residual regularity, stemming from the inherent properties of standard FSJ and ISRJ, still leaves a potential vulnerability exploitable by sophisticated ECCM algorithms.

**Breaking symmetry: controllable asymmetric deception via gated modulation**

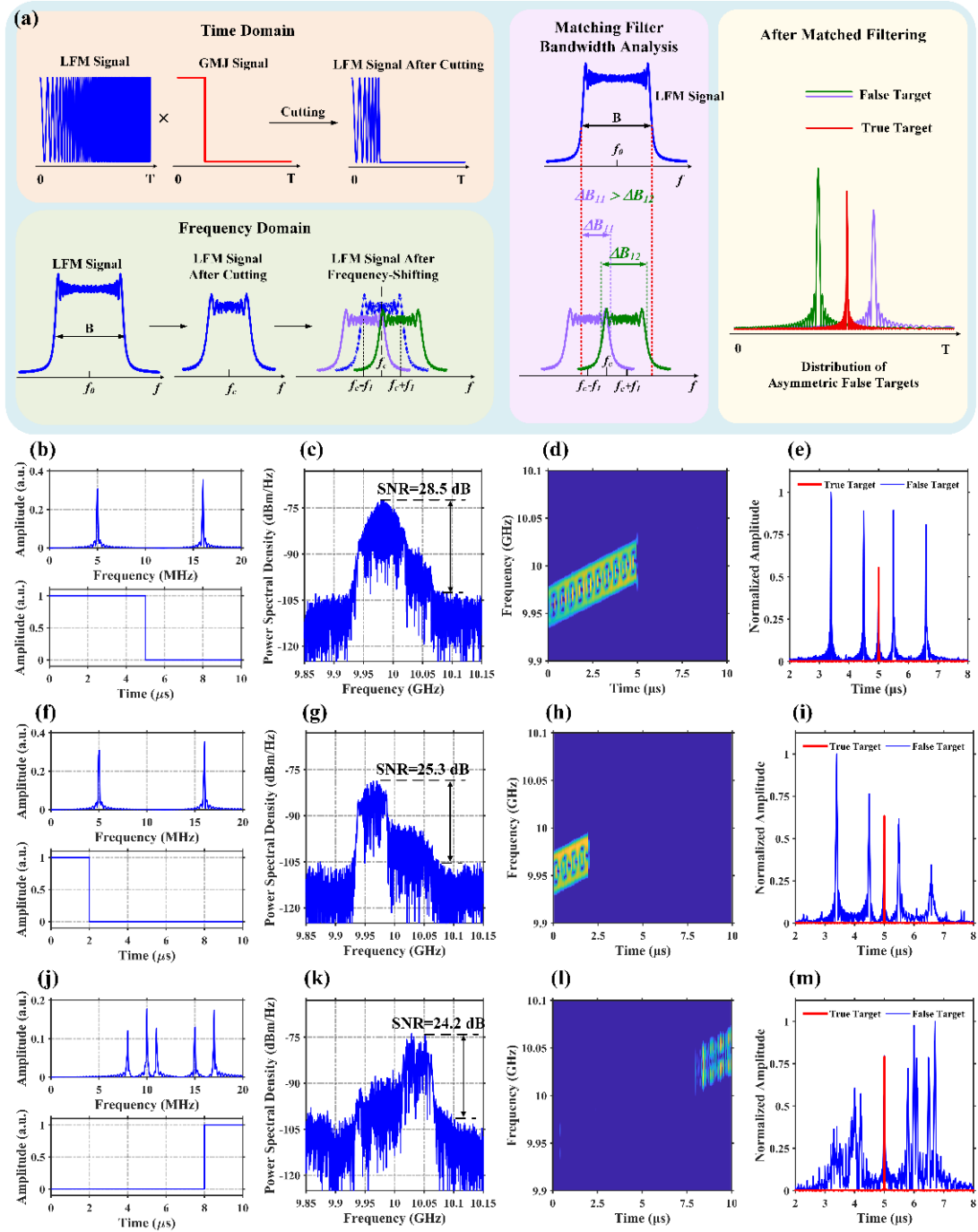

**Fig. 6 | Principle and experimental validation of on-chip asymmetric composite jamming (FSJ+GMJ). a** Schematic of the symmetry-breaking mechanism: An LFM signal is first frequency-shifted (multicast) and then truncated by an asymmetric time-domain gate. Due to time-frequency coupling, this truncation results in unequal effective bandwidths ($\Delta B_{11} > \Delta B_{12}$) for the sidebands, which translates into asymmetric amplitudes after matched filtering. **b–e** Experimental results for Experiment 1 (Gating pulse covers the first 50% of the LFM duration). **b** Electrical spectrum of the comb seed and time-domain waveform of the gating pulse. **c** Power spectral density. **d** Time-frequency analysis. **e** False-target distribution after matched filtering showing significant amplitude asymmetry. **f–i** Experimental results for Experiment 2 (Gating pulse covers the first 20%). **j–m** Experimental results for Experiment 3 (Gating pulse covers the last 20%).

To overcome the pervasive symmetry limitations in deceptive jamming, we propose and validate a novel composite jamming mode capable of actively generating asymmetric false-target distributions, namely Frequency-Shifting Jamming plus Gated Modulation (FSJ+GMJ). This mode leverages the unique time-frequency coupling property of LFM signals to break the inherent symmetry of traditional jamming. The principle is illustrated in Fig. 6a. First, the LFM signal is frequency-shifted to generate multiple sidebands. Then, an asymmetric time-domain gating pulse G(t) is applied to selectively truncate the signal. Since the instantaneous frequency of an LFM signal is linearly coupled to time, this time-domain truncation is equivalent to an asymmetric band-pass filtering in the frequency domain. As shown in the spectral diagrams in Fig. 6a, the gating operation causes the effective bandwidths of the upper and lower sidebands to differ ($\Delta B_{11} > \Delta B_{12}$). Consequently, after matched filtering, the pulse compression gain becomes unequal for different sidebands. Since the gain is proportional to the effective bandwidth, this process results in false targets with asymmetric amplitudes.

We experimentally validated this mechanism using the on-chip platform. The chip was configured similarly to the FSJ+ISRJ mode, but with MZM-3 driven by a single-shot gating pulse instead of a periodic train. Figure 6 presents the results under three distinct gating scenarios. In Experiment 1 (Fig. 6b–e), a gating pulse covering the first 50% of the LFM signal was applied. The matched filtering result (Fig. 6e) reveals a significant amplitude asymmetry with a measured centroid offset ($\Delta t_c$) of −0.14 μs, indicating a shift towards leading false targets. In Experiment 2 (Fig. 6f–i), shortening the gate to the first 20% dramatically enhanced this asymmetry and increased the negative offset to $\Delta t_c = −0.55$ μs. In Experiment 3 (Fig. 6j–m), by shifting the 20% gate to the end of the signal duration (80%–100%), we successfully reversed the direction of asymmetry and shifted the centroid to $\Delta t_c = +0.6$ μs, which favors lagging false targets.

This controllable asymmetry introduces a profound tactical advantage beyond simple deception. In traditional symmetric jamming, the power-weighted centroid of the false-target cluster often coincides with the true location of the jammer, making the platform vulnerable to "center-of-gravity" tracking algorithms. By contrast, our FSJ+GMJ mode enables the active displacement of the jamming centroid away from the true target. As demonstrated by the tunable $\Delta t_c$, which ranges from negative to positive values, we can strategically "pull" the radar's tracking gate toward either leading or lagging false targets. This capability not only enhances the deception success rate but also protects the jamming platform itself by decoupling its physical location from the perceived center of the interference source.

**SAR Imaging Simulation for Jamming Effectiveness Evaluation**

Finally, to evaluate the tactical effectiveness of our on-chip multi-mode jamming schemes in a realistic application context, we established a synthetic aperture radar (SAR) imaging simulation platform (simulation parameters are detailed in Supplementary Document). The results provide a direct visual comparison of jamming impacts on both point and extended targets.

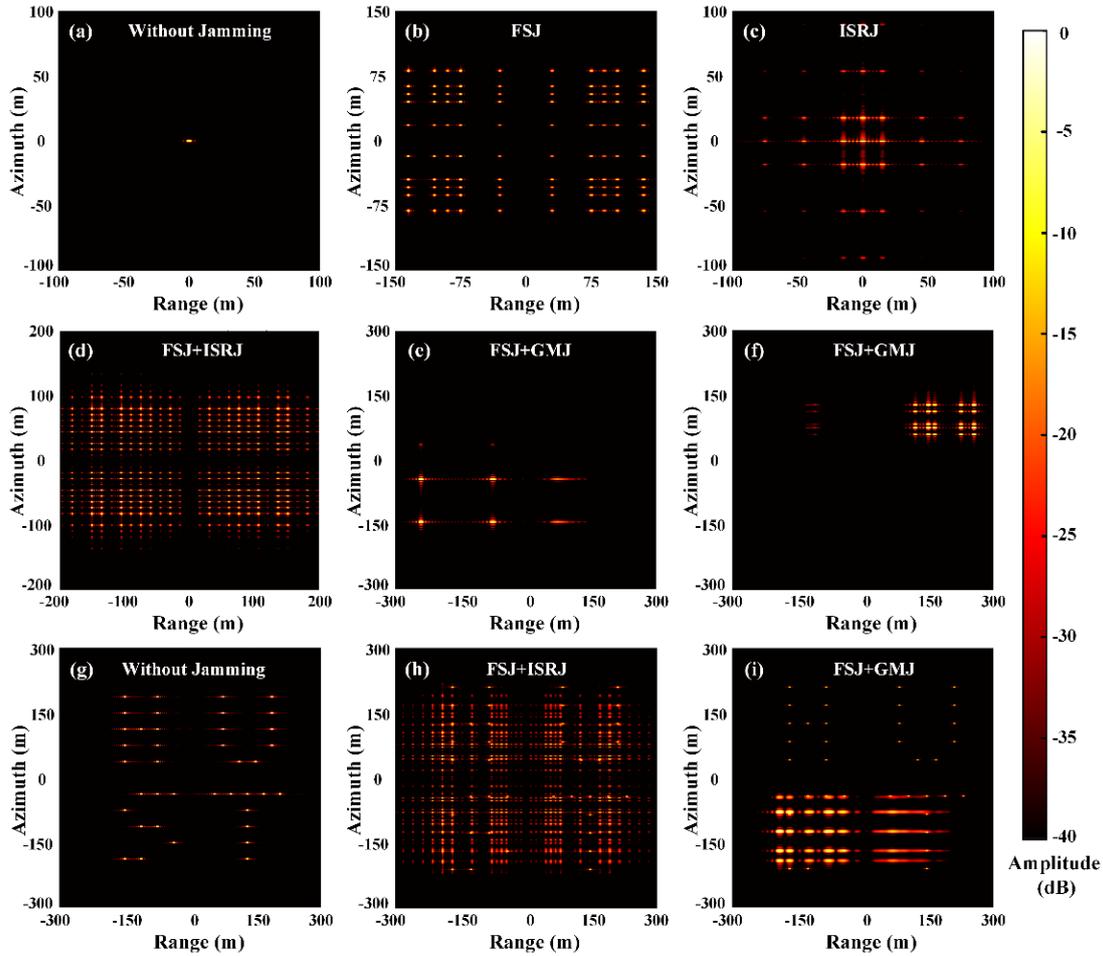

**Fig. 7 | SAR imaging simulation for jamming effectiveness evaluation. a–f** Imaging results for a point target. **a** Reference image without jamming. **b** FSJ mode showing symmetric false targets. **c** ISRJ mode showing a regular grid of false targets. **d** Compound Jamming (FSJ+ISRJ) creating a dense but symmetric jamming texture. **e** Asymmetric Compound Jamming (FSJ+GMJ) configured for leading false targets (negative centroid offset). **f** Asymmetric Compound Jamming (FSJ+GMJ) configured for lagging false targets (positive centroid offset). **g–i** Imaging results for an extended target ("HUST"). **g** Reference image without jamming. **h** Compound Jamming (FSJ+ISRJ) obscuring the target but maintaining a symmetric energy distribution centered on the jammer. **i** Asymmetric Compound Jamming (FSJ+GMJ) distorting the target while shifting the energy centroid away from the true location, demonstrating effective range-gate pull-off.

In the baseline scenario without jamming, both the point target (Fig. 7a) and the extended target "HUST" (Fig. 7h) are clearly imaged. When the chip's foundational modes are activated, FSJ (Fig. 7b) and ISRJ (Fig. 7c) generate arrays of false targets. However, these distributions exhibit highly predictable symmetry and regularity, which can be easily identified by modern ECCM algorithms. Advancing to the first composite mode (FSJ+ISRJ), the system generates a dense jamming grid that successfully obscures the extended target (Fig. 7d and Fig. 7h). However, a critical tactical vulnerability persists: the jamming energy remains symmetrically distributed around the jammer's true location (the coordinate origin). Consequently, while effectively obscuring the target's structural details, the zero-mean centroid of the jamming signal provides a deterministic reference for enemy localization algorithms, thereby compromising the platform's survivability.

In stark contrast, our asymmetric FSJ+GMJ mode demonstrates a sophisticated tactical advantage, achieving simultaneous deception and self-preservation. For the point target, we can flexibly steer the false-target cluster to either lead (Fig. 7e) or lag (Fig. 7f) the real object by reconfiguring the gating pulse. When applied to the extended target (Fig. 7i), this mode not only obliterates the target's structural features but also concentrates the jamming energy into a localized region significantly offset from the center. This strategic decoupling of the jamming centroid from the physical location provides a twofold benefit: it effectively pulls the radar's tracking gate away from the real target (deception) while simultaneously misleading counter-analysis algorithms attempting to locate the interference source (anti-localization). These results compellingly validate the superior survivability and deception capabilities of our monolithic asymmetric jamming solution.

# Discussion

**Table 1. Performance Comparison with Existing Radar Jamming Schemes**

| Ref. | Integrated (Y/N) | Working Frequency | Operating Bandwidth (GHz) | Number of Jamming Modes | Jamming Format | Asymmetry of False Target Distribution (Y/N) |
|---|---|---|---|---|---|---|
| [10] | N | C to X | ~8 | 1 | ISRJ | N |
| [23] | N | C to Ka | ~36 | 1 | FSJ | N |
| [16] | N | C to Ku | ~14 | 2 | CSMJ, ISRJ | N |
| [22] | N | S to K | 18 | 1 | FSJ | Y |
| [26] | N | X to Ka | 20 | 3 | ISRJ, CSMJ, ISRJ+CSMJ | N |
| [21] | N | C to Ka | ~32 | 3 | FSJ, ISRJ, FSJ+ISRJ | N |
| [27] | N | L to X | 9 | 1 | PMJ | N |
| [28] | N | C to Ku | >20 | 1 | ISRRJ | Y |
| **This Work** | **Y** | **C to Ku** | **>20** | **4** | **FSJ, ISRJ, FSJ+ISRJ, FSJ+GMJ** | **Y** |

**Note:** ISRJ: Interrupted-Sampling Repeater Jamming; FSJ: Frequency-Shifting Jamming; CSMJ: Comb Spectrum Modulation Jamming; PMJ: Phase-Modulated Jamming; ISRRJ: Interrupted-Sampling Repetitive Repeater Jamming; GMJ: Gated-modulation jamming.

We have designed, fabricated, and validated a monolithic silicon photonic integrated circuit that addresses the critical challenges of integration density, operating bandwidth, and jamming sophistication in modern electronic warfare. By monolithically integrating a high-speed modulator array and a photodetector within a compact 2×5 mm² footprint, our solution achieves a transformative improvement in system stability and form factor compared to discrete fiber-optic systems. As comprehensively summarized in Table 1, our work stands out as the only solution that simultaneously achieves monolithic integration, an operating bandwidth exceeding 20 GHz, and the capability to generate asymmetric false targets. The platform's wideband operation fundamentally circumvents the electronic bottlenecks of traditional DRFMs, while its ability to be electrically programmed into four distinct jamming modes—ranging from foundational FSJ and ISRJ to complex composite variations—demonstrates a level of reconfigurability that surpasses single-function photonic solutions.

Beyond platform-level advancements, the central scientific contribution of this work lies in the first demonstration of controllable asymmetric deception on an integrated photonic platform. The symmetric and regular false-target distributions generated by traditional jamming methods represent a fundamental vulnerability, as modern ECCM systems can readily identify and suppress these deterministic signatures. Our proposed gated-modulation scheme (FSJ+GMJ) breaks this inherent symmetry by exploiting the time-frequency coupling of chirped radar signals. This breakthrough yields a twofold tactical advantage: First, the generation of irregular waveforms significantly enhances counter-detection capabilities against template-matching algorithms. Second, and more critically, the controllable asymmetry enables the active displacement of the jamming centroid. As demonstrated in the SAR imaging results, our scheme strategically decouples the perceived signal source from the platform's physical location. This capability not only effectively deceives the radar's tracking logic but also maximizes platform survivability by misleading counter-analysis attempts.

While this work marks a significant milestone in integrated photonic jamming, the path toward a fully self-contained system-on-chip involves further integration steps. Currently, the system relies on an off-chip laser source and electrical power amplifier. Future iterations could leverage heterogeneous integration to co-package III-V gain materials (for lasers and SOAs) and CMOS driver circuits directly with the silicon photonic core, thereby minimizing system volume and power consumption. Furthermore, the versatile reconfigurability demonstrated by our chip provides an ideal hardware interface for synergy with AI-driven cognitive electronic warfare engines. By combining real-time spectral sensing with the agile mode-switching capability of our processor, future systems could autonomously adapt jamming strategies to dynamic radar threats, paving the way for truly intelligent electromagnetic dominance.

# Methods

**Device fabrication and packaging**

The photonic integrated chip was fabricated at a commercial CMOS-compatible silicon photonics foundry (CUMEC). The device is built on a standard silicon-on-insulator (SOI) wafer featuring a 220-nm-thick top silicon layer and a 2-μm buried oxide (BOX) layer. Active on-chip components, including the single-drive series push-pull Mach-Zehnder modulators (SPP-MZMs) and high-speed germanium-on-silicon (Ge-on-Si) photodetectors, were realized through foundry-standard doping and deposition processes. For electronic interfacing, the die was wire-bonded onto a custom-designed high-speed printed circuit board (PCB). Optical coupling was achieved using a high-precision fiber array (FA) aligned to the edge couplers on the chip facets. DC bias voltages, required for MZM thermal phase tuning and PD reverse biasing, were supplied via DC electrical interfaces on the PCB.

**Experimental Setup**

The experimental validation was conducted using the setup illustrated in Fig. 1c. A continuous-wave (CW) optical carrier was provided by a tunable laser source (KG-TLS) operating at 1550 nm with an output power of 16 dBm. The light was routed through a polarization controller (PC) to optimize coupling into the photonic chip. The electrical drive signals were synthesized by two arbitrary waveform generators (AWGs): a high-speed AWG (Tektronix AWG70001A) generated the wideband linear frequency-modulated (LFM) radar signals, while a second AWG (Ceyear AWG1652B) generated the specific jamming seed signals (electrical comb, periodic rectangular pulse, or asymmetric gating pulses) corresponding to the selected jamming mode. These RF signals were delivered to the respective MZM driving ports via high-frequency coaxial cables. A multi-channel voltage source (MCVS6400-A) provided the necessary DC bias points for the modulators. The generated jamming signal output from the on-chip PD was amplified by a medium power amplifier (IWZA-010200G23) and subsequently digitized for offline analysis using a high-bandwidth real-time oscilloscope (Tektronix DPO72004B).

**Data Processing and Analysis**

All time-domain waveform data captured by the oscilloscope were exported and processed using MATLAB.

**Matched Filtering:** To evaluate the jamming effectiveness and pulse compression characteristics, the captured jamming signal, $s_j(t)$, was digitally processed using a matched filter. This operation was implemented by calculating the cross-correlation between $s_j(t)$ and the time-reversed complex conjugate of the ideal reference radar signal, $s_{ref}^*(t)$.

**Centroid Offset Calculation:** To quantify the tactical deception capability of the asymmetric jamming mode, we calculated the power-weighted centroid offset of the generated false-target cluster. First, a peak detection algorithm extracted the relative delay $t_i$ and peak power $P_i$ of each discrete false target from the matched-filter output. The power-weighted centroid $\bar{t}$ of the cluster was calculated as: $\bar{t} = \sum(P_i \cdot t_i)/\sum P_i$.

Finally, the centroid offset $\Delta t_c$ is defined as the difference between the calculated centroid $\bar{t}$ and the true target's position $t_{true}$: $\Delta t_c = |\bar{t} - t_{true}|$. Note that the sign of $\Delta t_c$ indicates the direction of the deception, with negative and positive values corresponding to leading and lagging false targets, respectively.

## Supplementary information

Supplementary Information is linked to the online version of the paper.

## Author contributions

## Acknowledgments

This work was supported by the Natural Science Foundation of Hubei Province of China (2024AFA016, 2024AFB612), National Natural Science Foundation of China (NSFC) (62475085, 62435006), Wuhan Key R&D Program (2025050602030052).

## Data Availability

The data that support the findings of this study are available from the corresponding authors on reasonable request.

## Competing interests

The authors declare no competing interests.